# *Plasmonic nonvolatile memory crossbar arrays for artificial neural networks*

Jacek Gosciniak

*Independent Researcher, 90-132 Lodz, Poland*
*Email: jeckug10@yahoo.com.sg*

**Abstract**

Here it is proposed a three-dimensional plasmonic nonvolatile memory crossbar arrays that can ensure a dual-mode operation in electrical and optical domains. This can be realized through plasmonics that serves as a bridge between photonics and electronics as the metal electrode is part of the waveguide. The proposed arrangement is based on low-loss long-range dielectric-loaded surface plasmon polariton waveguide where a metal stripe is placed between a buffer layer and ridge. To achieve a dual-mode operation the materials were defined that can provide both electrical and optical modulation functionality.

**Introduction**

Neuromorphic plasmonics combines the advantages of plasmonics [1] and neuromorphic architecture [2, 3, 4, 5] to build systems with high interconnectivity, high efficiency and high information density.

It is based on plasmonics that combines the advantages of photonics (optics) and electronics to create low cost, ultrafast and low footprint signal processing units for the next generation neural networks [6, 7]. Until today, most of a develop neuromorphic networks utilize electronics as main building blocks for future neuromorphic computing [5, 8, 9, 10, 11, 12]. However, they suffer from a low bandwidth and low energy efficiency [4, 13].

In a contrary, photonics is a very proposing platform for neural networks as it provides tremendous bandwidth, multiplexing capabilities and the possibility of co-integration with electronics [7, 9, 14, 15, 16, 17, 18]. Additionally, it enables high-speed operation, high power efficiency, and law latency what can bring additional benefits for brain-inspired approaches.

As the main computational burden in neural networks lies in the interconnectivity, photonic systems can address this problem. Interconnectivity can be significantly boost by waveguides carrying many signals at the same time. Furthermore, photonics that is characterized by low-energy operations can reduce the computational burden of performing linear functions such as weighted addition. As a result, a tremendous progress can be made if we notice that a neuromorphic photonics can potentially operate 6-8 orders of magnitude faster than neuromorphic electronics [11, 12, 19, 20, 21, 22, 23, 24] (Fig. 1).

However, apart from scalability and system stability the main limitations of neuromorphic photonics lie in interfacing with electronic systems [2]. This can be overcome by plasmonics where metal electrode is part of the plasmonic waveguide and can, simultaneously, serve as a base for electronic systems. In consequence, the progress in neural networks can be hugely boosted (Fig. 1).

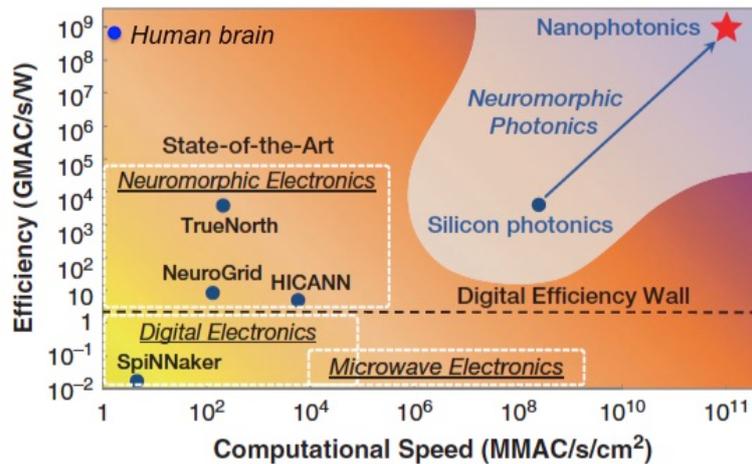

**Figure 1.** Comparison of various neuromorphic hardware platforms. The regions highlighted in the graph are approximate based on qualitative trade-offs of each technology [11].

Plasmonics is a very promising approach for realization of artificial neural networks - it serves as a bridge between photonics and electronics where a metal electrode(s) is part of the waveguide [1, 25, 26, 27, 28, 29, 30, 31, 32, 33, 34]. Thus, it can ensure a dual-mode operation as it provides both electrical and optical modulation functionality. As plasmonics is usually characterized by relatively high losses of metals at optical frequencies, it can be combined with photonics where photonics can be served as a low-loss delivery system and plasmonics can provide both strong light-matter interaction and reduced device footprint. However, some plasmonic waveguide arrangements can provide low absorption losses in metal and good mode confinement [25, 26, 27, 28, 29, 30], thus they can serves as a platform for artificial neural networks (ANN).

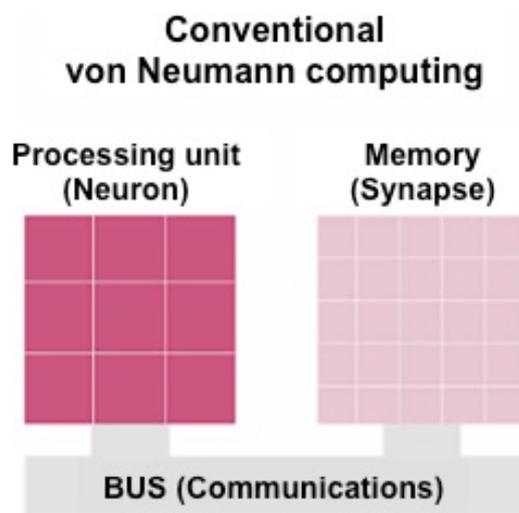

**Figure 2.** Conventional von Neumann computer architecture [4].

## Artificial neural networks

Conventional microprocessors follow the so-called von Neumann architecture where the machine instructions and data are stored in memory and share a central communication channel to a processing unit, thus data are continually moved back and forth between memory and the processor [4] (Fig. 2). This significantly limits the speed and energy efficiency, and the performance mismatch between this two units leads to considerable latency. It was showed that real-time applications such as audio recognition and hand-tracking services consume more than half of the total energy only for moving and storing data without performing any computations [4].
Furthermore, as the channel length and thickness of gate dielectric of a transistor approach the scaling limit, the leakage currents start to be a problem [10].

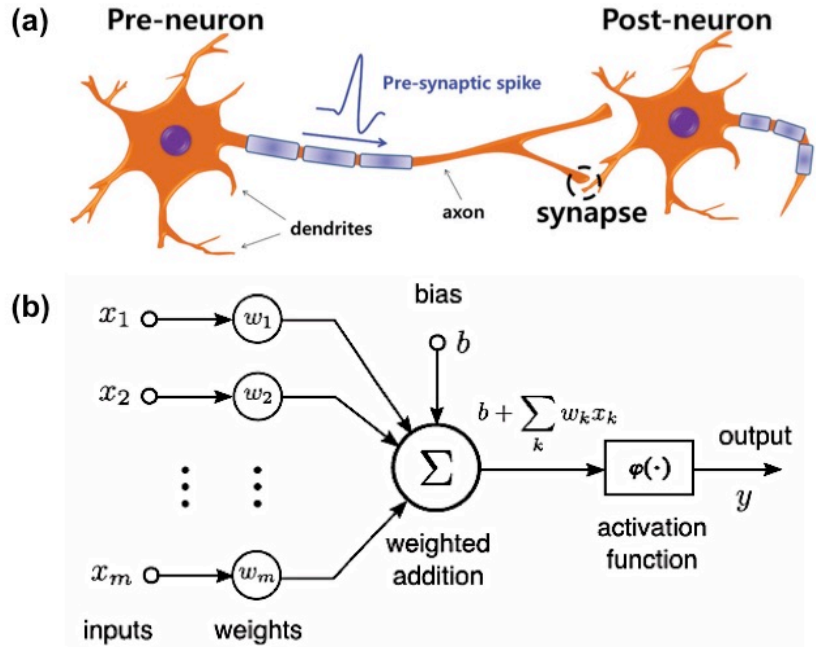

**Figure 3.** (a) Schematic illustration of biological neurons and synapse. Signals are transferred from pre-to post-synaptic neuron through a synapse [35]. (b) Nonlinear model of a neuron where each signal is weighted by the corresponding synaptic connection. The weighted signals reach the neuron, where they are summed together before the nonlinear activation function is applied [12].

Most of the available today ANNs utilize more energy efficient algorithms while still being realized using conventional von Neumann architecture resulting in processing bottlenecks and high power consumption [4]. To enhance processing capabilities it will be highly desired to fuse together the two main tasks of processing and memory. In recent years the search for new computing approaches has been boosted by the advent of memristive devices that can both process and store data simultaneously [3, 10, 36, 37]. This approach provides a range of attractive properties including intrinsic parallelism, learning, adaptive capabilities and the simultaneous execution of processing and storage [3, 10, 37, 38]. It is inspired by the human brain that is characterized the co-location of logic and memory, hyper-connectivity and parallel processing.

Compared to standard architecture that utilizes digital 0's and 1's, the neural networks represent information in analog signals [11, 12]. Compared to a sequential set of instructions, neurons process data in parallel and are programmed by the connections between them [37, 11].

The ANNs are inspired by neuroscience, i.e., there are based on the structure of human brain where the parallel and energy efficient data processing is realized by interconnection of ~$10^{11}$ neurons and ~$10^{15}$ synapses that perform memory and learning [13]. The basic function of neuron is to receive and transmit complex spatio-temporal information. It consists of a soma, dendrites, and axons. When dendrites receive a signal from the axon terminal of the pre-synaptic neuron, the some integrates the signal and generates action potentials only when the input spikes exceed a threshold value. After action potentials are generated, the signals are transmitted along axon. When the signals arrive at the end of the axon, they are delivered through the synapse to the dendrite of the post-synaptic neuron [11, 13]. Neurons communicate with each other through synapses that are responsible for processing and memorizing the information about signals from neurons (Fig. 3a). The connection strength between two neurons is referred to as the synaptic weight and is dependent on the recent stimulation history of the synapse. The activity-dependent change in synaptic weight is due to synaptic plasticity, i.e., the ability of synaptic connections to strengthen or weaken over time depending on external stimulation, that can increases under continuous and repeated stimulation of pre-synaptic neurons. Synaptic plasticity can be classified as short-term plasticity (STP) or long-term plasticity (LTP) according to the duration for which it is retained. While in STP the synaptic weights are temporarily changed, the LTP is responsible for a long-term change in synaptic weights, thus, it play an important function in learning and memory [10].

ANNs attempt to mimic the natural processing capabilities in the brain (Fig.3b). The main three elements that constitute neural networks are: a set of nonlinear nodes (neurons), configurable interconnection (network), and information representation (coding scheme) [11]. The network consists of a weighted directed graph, in which connections are called synapses (Fig. 3b). The input into particular neurons is a linear combination, i.e., weight addition, of the output of other neurons connected to it. Then, the signal from the particular neurons is integrated and combined and produces a nonlinear response represented by an activation function. The weighting is therefore represented as a real number, and the interconnection network can be expressed as a matrix.

Early attempts in neuromorphic computing architectures such as TrueNorth chip from IBM or SpiNNaker from EU have relied mostly on silicon-based CMOS materials [11] (Fig. 1). However, CMOS chips suffer from energy inefficient operations as they are based on volatile random access memory. Thus, searching for a non-volatile memory as a foundation for neuromorphic computing became essential as it can store the synaptic weights using its multiple conductance states and performs matrix multiplication and addition on-chip without accessing external memory.

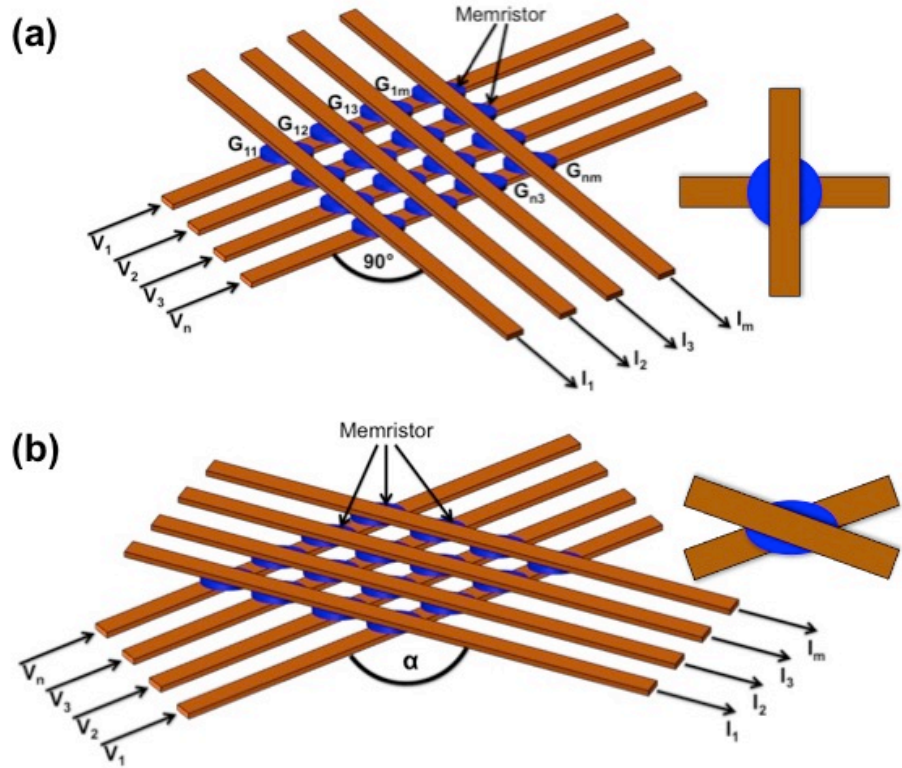

**Figure 4.** Memristor crossbar array with corresponding applied (read) voltages $V_n$ and conductance of memristor devices $G_{nm}$ resulting in sensed currents $I_m$ for (a) 90° and (b) variable angle α between two adjacent layers.

## Memristors

Compared to the digital computers that operate based on binary components, the brain-inspired ANNs utilizes multilevel logic components called synapses that do not consume energy (nonvolatile) while in a passive state. Memristors, owing to its nonvolatile properties, can be used as an artificial synapse to emulate biological synapses [3, 10, 35, 37, 38].

Memristors [39, 40, 41], also called the resistance switches, are the essential building blocks of future artificial neural networks whose internal states are dependent on the history of the current [3, 10, 35, 36, 37, 38, 39, 40]. They can be organized into large arrays or/and stacked three-dimensionally (3D) where the computing results can be stored locally as conductance in each single, non-volatile memristor during computing (Fig. 4).

Memristors can be integrated at the junctions of crossbar arrays to represent the weight of synapses as conductance at each cross-point [3, 10, 36, 37, 38, 42, 43]. Thus, a density is expected to overcome the scaling limit of CMOS-based computing. Furthermore, memristive neural networks are promised to provide orders of magnitude higher speed-energy efficiency product then traditional CMOS hardware platform. It shows over 3 orders of magnitude energy-per-compute reductions defined by multiply-and-accumulate (MAC) function [12].

The MAC is a very important parameter that can be used for evaluation of any new hardware technologies for neutral networks [19]. The key metric is the number of processed MAC's per unit time and unit power, i.e., MAC/s/W, which defines the efficiency and performance of the system. It shows huge improvement over previous technologies. However, no improvement in terms of the computational speed (MAC/s/cm$^2$) was observed compared to regular von

Neumann systems due to the high RC delay of electronic circuits (Fig. 1).
Memristive crossbar array is very efficient at vector-matrix multiplication (VMM) and enables to perform MAC operations in-place, i.e., in-memory, and in the analog domain by exploiting Ohm's law (multiply operation) and Kirchhoff's law (accumulate operation) [3, 10, 36, 37]. When the input signal is applied as voltage pulses across a conductor, to the rows of the crossbar, the resulting current, that is the multiplication of the voltage and the conductance, is collected at the columns of the crossbar. Combined current from multiple individual conductors leads to the accumulated current. Each memristive device represents a synaptic weight while the memristive crossbar arrays represent weights in ANNs as conductance at each cross-point. This approach allows direct computing of data-intensive task both in memory and in parallel in a single step and highly reduces the extensive data movement required by conventional digital systems based on the von-Neumann architecture. Furthermore, it can be integrated with CMOS technology to ensure huge efficiency improvement over existing technology. Thus, it can ensure non-volatility, long retention, high endurance, short switching time and compatibility with standard silicon processing technology that are highly required for an artificial neutral networks.

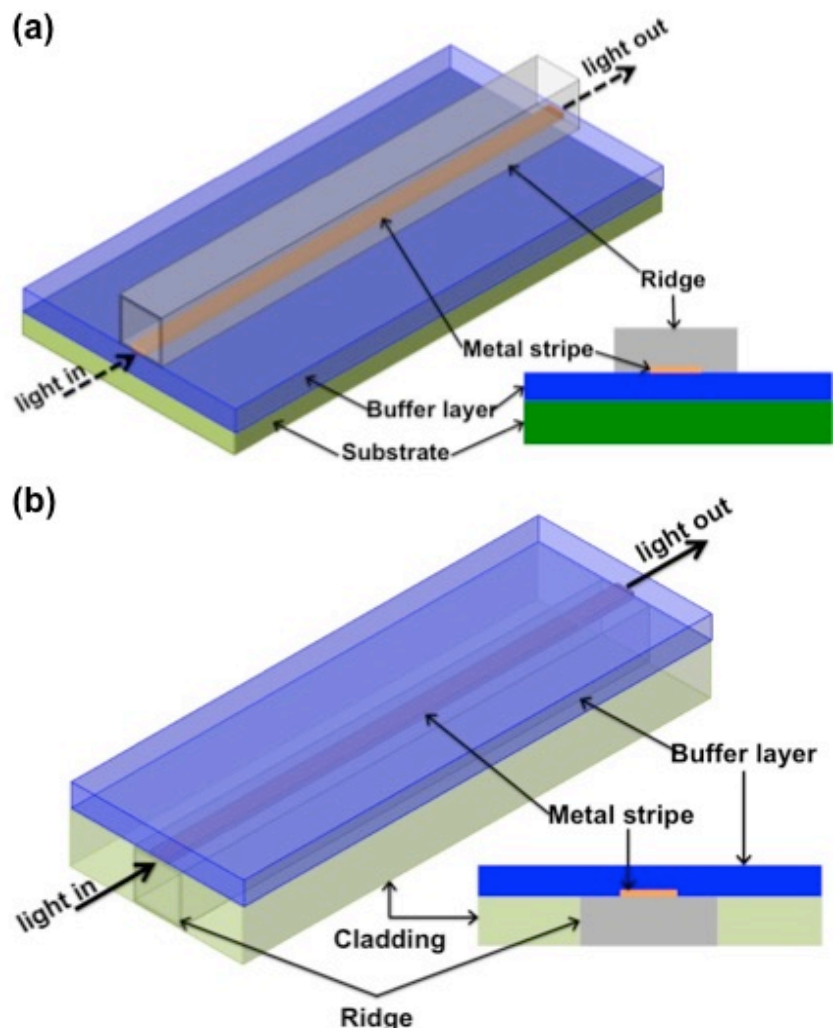

**Figure 5.** Schematic of the proposed LR-DSLPP plasmonic waveguide arrangement that serves as a main building component for a plasmonic neural network in (a) "normal" and (b) "inverse" designs.

## LR-DLSPP waveguide platform

The proposed here ANN is based on a long-range dielectric-loaded surface plasmon polariton (LR-DLSPP) waveguide configuration with a plasmonic mode guided by a metal stripe that serves simultaneously as the metal electrode [26, 27, 28, 29, 30, 34] (Fig. 5). In this arrangement, the metal stripes can serve as a frame for memristors arranged at the crossbar arrays as presented in Fig. 4.

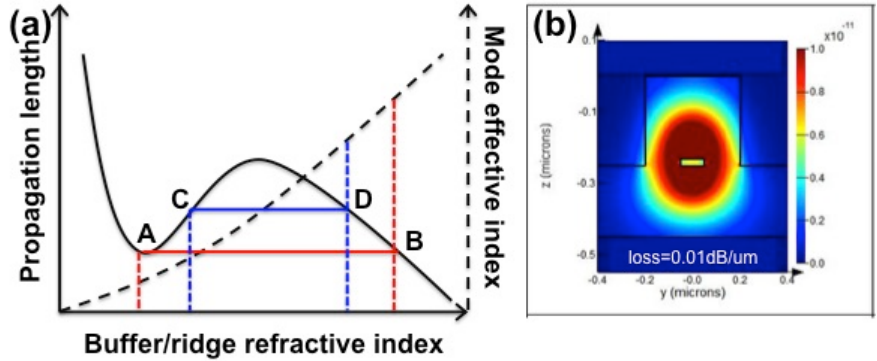

**Figure 6**. (a) Propagation length (solid line) and mode effective index (dashed line) as a function of buffer layer / ridge refractive index and (b) mode effective index and calculated losses for LR-DLSPP waveguide with Au stripe.

For LR-DLSPP waveguide when the mode effective index below a metal stripe is equal or close to the mode effective index above a metal stripe a balance is reached (Fig. 6). At this point the longitudinal component of the electric field in the metal is minimized, thus lowering the absorption losses that results in long propagation length of the mode (Fig. 6b). As the refractive index of the buffer or ridge increase/decrease, the unbalance in the mode fields appears that affects the propagation length of the mode and mode becomes bound the to metal stripe.

However, as it can be seen from Fig. 6a it is possible to engineer the LR-DLSPP waveguide in such a way that the change in buffer layer or/and ridge refractive index under heating, electrical pulses or optical pulses will not effect a propagation length, i.e., absorption losses, while it will effect the mode effective index of the LR-DLSPP waveguide (red and blue lines in Fig. 6a). Thus, the transmitted light can be tuned not via absorption but through phase change. As a result, the absorption losses under a switching approach can be avoided or highly reduced.

Depending on the requirements, LR-DLSPP can be integrated in ANN with the "normal" design with the ridge on top of the metal stripe (Fig. 5a) or in "inverse" design with the ridge below a metal stripe (Fig. 5b). Thus, it can provide an additional flexibility in a design and enable an easy integration in the multilayer arrangements. The examples of the two layers integrated components are showed in Fig. 7 for two opposite arrangement stacks and different size of a memory material that can be farther integrated in multilayer arrangements.

The proposed plasmonic arrangement enables integration of photonics and electronics in one component, thus it can ensure a dual-mode operation – electrical and optical.

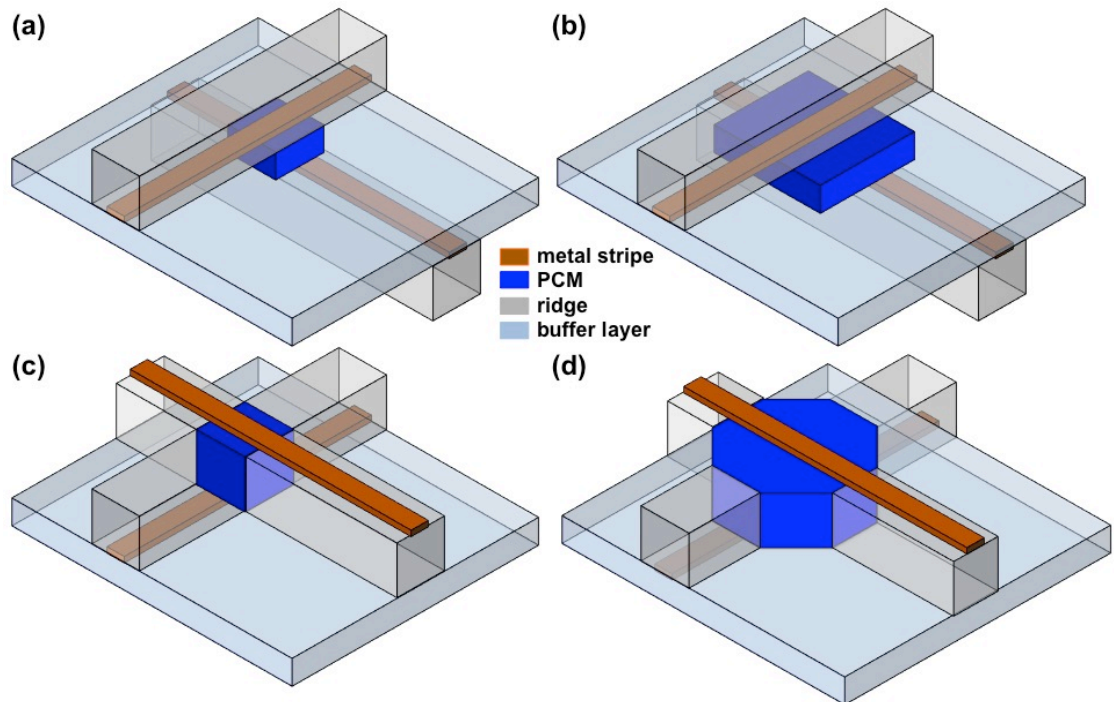

**Figure 7**. Memristors organized in plasmonic waveguide arrangements in (a, b) "inverse" and (c, d) "normal" design.

## Emerging non-volatile memories

Among many technologies that have emerged in the last few years, most of the attention is paid to a resistive random access memory (RRAM) [44, 45, 46], phase-change memory (PCM) [6, 7, 8, 9, 14, 15, 16, 17, 18, 47, 48, 49], ferroelectric tunnel junctions (FTJs) [50, 51, 52, 53, 54, 55] and 2D materials [2, 56, 57, 58].

RRAM technology consists of a simple metal-insulator-metal (MIM) layer structure where the operation principle is based on the switching of the electrical resistance between low and high resistance state. Recently, a promising approach focused on conductive bridge (CB) cells where the resistance switching is achieved through the formation or annihilation of the nanoscale metal filament under the electrical voltage or light illumination [44, 45, 46]. Consequently, this CB-RRAM can be electrically and/or optically controlled and can provide some advantages such as low switching electrical power, retention, endurance and high operation speed.

In PCM the resistance depends on the crystal structure of the chalcogenide materials where the two phases can be reversibly changed by first melting the solid-state chalcogenides into a glassy state and then controlling the time required for the ions to be rearranged [5, 17]. To generate heat, a confined electrode serving as a heater is used to maximize the current density by reducing the region in which current flows. Under an electrical pulse the Joule heating is induced that cause melting of material near the electrode.

For a short pulse, the amorphous, disorder state is achieved that is described by a high resistance state (HRS), known as a resent process. Meanwhile, when sufficient time to relocate the ions to a thermodynamically stable position is provided during the molten state, the crystalline state can be formed to obtain a low resistance state (LRS), known as a set process [4].

The ferroelectric technology is based on the ferroelectric materials that exhibit a spontaneous electric polarization in the absence of an external electric field. However, under an external electric field the polarization can be switched showing typical hysteretic characteristics [59]. Ferroelectrics are very promising materials for nonvolatile memories and artificial synapses [52, 53, 54, 60]. In recent years a new class of ferroelectric materials emerged, so called van der Waals (vdW) ferroelectric materials that offer additional merits such as bandgap tunability, mechanical flexibility and high carrier mobility [50, 53, 54, 55]. They show wide bandgap ranging from 0.18 eV for SnTe, to 1.36 eV for α-$In_2Se_3$ [50].

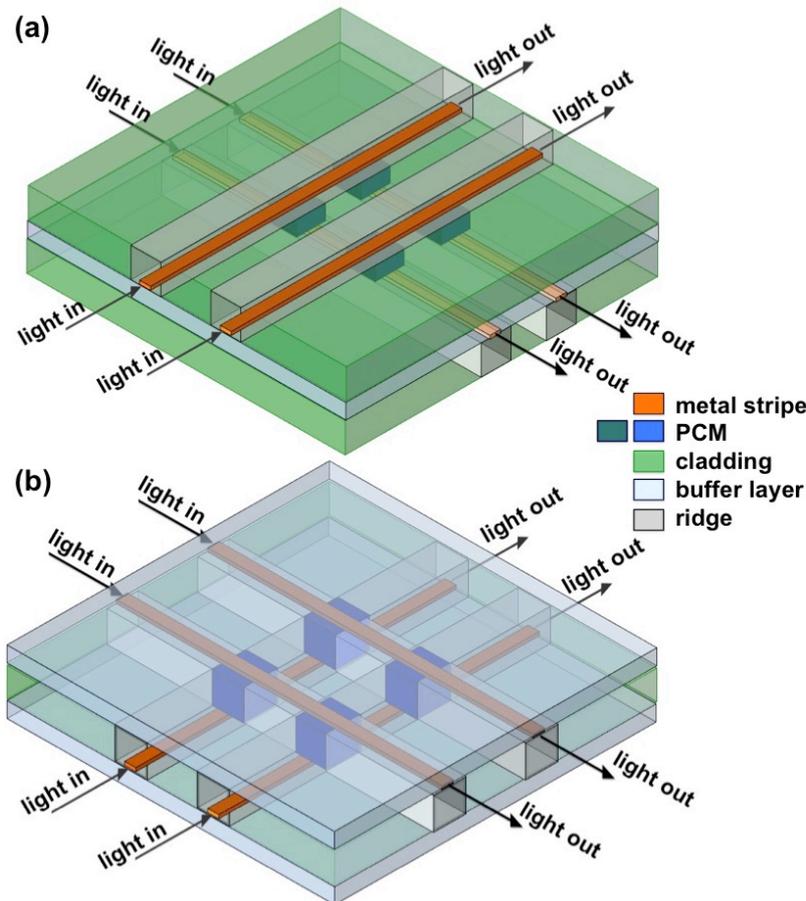

**Figure 8.** Sketch of LR-DLSPP plasmonic waveguide crossing arrays organized in (a) "inverse" and (b) "normal" designs.

Two-dimenssional (2D) materials represent a very interesting class of materials that show excellent electronic properties, electrostatic doping, gate- and optically-tunable photoresponse, and high thermal stability [2]. Stacking distinct 2D materials on top of each other enables creation of diverse vdW heterostructures with different combination and stacking orders exhibiting intriguing electrical and optical properties beyond those of individual 2D materials [61]. Recently, 2D-based memristive device was reported to exhibit endurance of $10^7$ at room temperature and stable switching performance in high operating temperature of 340 °C [61].

**Memristive plasmonic crossbar arrays**
In Fig. 8, the LR-DLSPP plasmonic crossbar arrays are presented in "normal" and "inverse" arrangement that are based on the LR-DLSPP waveguide design (Fig. 5) in which the metal stripes create the metallic crossbar array as presented in Fig. 4. Here, the PCM is used as a nonvolatilte switching material.

*a. PCM*

PCMs are perfect candidates for dual-mode operation as they provide both electrical and optical modulation functionality [6, 7, 9, 15, 17, 48, 49]. Thus, the reversible phase transition of PCMs on Photonic Integrated Circuits (PICs) can be performed either by optical or electrical heating. On-chip optical heating by optical pulses suffers however from the relatively transparent amorphous state in the recrystallization process [9] (Fig. 9). It can pose a serious problem in a photonic waveguides.

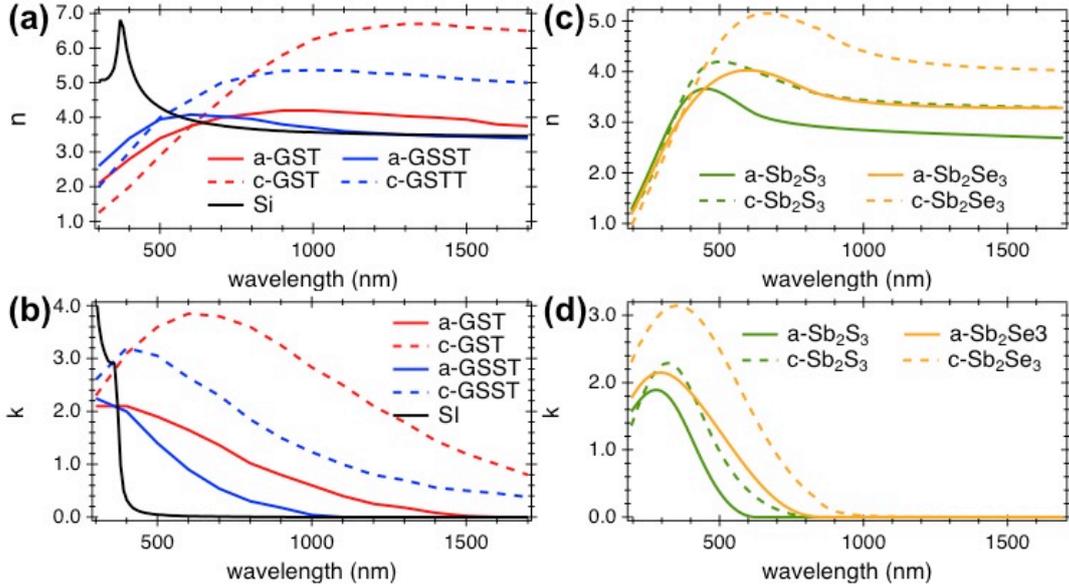

**Figure 9**. (a, c) Real and (b, d) imaginary parts of refractive index of the amorphous and crystalline phased of $Ge_2Sb_2Te_5$ (GST), $Ge_2Sb_2Se_4Te_1$ (GSST) [62], $Sb_2S_3$, and $Sb_2Se_3$ [63] from the visible range to near-infrared. The results were compared with Si.

However, in the proposed LR-DLSPP waveguide arrangement the optical heating can be highly enhanced. In the absence of the optical pulse the mode effective indices on both sides of the metal stripe are in balance, thus the absorption losses in metal are minimalized. Under absorption of light by the PCM the balance is disturbed and, in consequence, the absorption in metal arise. As a result, the metal stripe is heated and can work as an additional heater for the PCMs [31, 32].

Simultaneously, for an electrical heating, the change of PCM phase under an applied voltage is realized under a heating of PCM. Usually, it is done via metals [6], transparent conductive oxides [18], doped silicon [9] or graphene [49].

Doped silicon is a good choice for PCM integration with the silicon-on-insulator platform, however it is challenging to apply it to $Si_3N_4$-based devices or other non-silicon waveguide platform [9]. In comparison, graphene similarly to TCO suffer from optical losses in the infrared due to free carrier absorption [18, 49].

In traditional photonic devices metals introduce significant optical losses in waveguide components, thus they are not the first choice materials for a heater. However, in LR-DLSPP waveguide arrangement the metal is part of the plasmonic waveguide that ensure a reduction of the absorption losses in metal through a special design. Thus, it can serve as heater and provide a reversible switching of PCM while keeping the absorption losses at the minimum.

Here, the combined waveguide-integrated LR-DLSPP with PCM creates an electro-optical memory cell that is fully addressable in both electrical and optical domains. The strong field confinement and compact dimensions enable both electrical and optical nonvolatile switching of PCM and allows for full mixed-mode operation of a PCM memory cell. It can be achieved as the PCM-based devices are characterized by the high contrast in both the electrical (resistivity) and optical (refractive index) properties of PCMs between their amorphous and crystalline states (Fig. 9). In terms of the optical properties it means that optical pulses can both switch the refractive index of PCM and then it can be used to readout the resulting state. For chalcogenide PCM such as GST, the resulting state can be retained for more than ten years under ambient conditions and can endurance up to $10^{15}$ cycles while simultaneously it can be reversibly switched on a sub-nanosecond timescale [47].

The switching energy and speed in PCM depends mostly on the heating of the PCM unit cell thus the reduction in switching time and energy can be achieved by reduction of the unit-cell volume and/or magnification of the light-matter interaction. It can be achieved by plasmonics that ensures both a field enhancement and reduced device footprint.

Most of the previously reported nonvolatile photonic memories were realized by cladding waveguides with PCMs [16, 20, 48]. Here however, the PCM is integrated part of the waveguide with the PCM implemented in the cross of the metal stripes either in the buffer layer or in the ridge depending on the configuration (Fig. 7). Changing the state of PCM under a current flow either in the buffer layer or in the ridge from amorphous to crystalline states varies the mode effective index either above or below metal stripe and pushes the mode either to the ridge or to the buffer layer. Under a current flow a balance between mode effective index below a metal stripe and above a metal stripe is disturbed and plasmonic mode is pushed on one side of the metal stripe. In consequence, it affects the transmission of the light through the waveguide. Thus, the proposed configuration offers non-volatile weights that can be reconfigured with optical or electrical signals. Furthermore, this process of setting the weights is also reversible.

Presented configuration can be used for both optical and electrical programming of the phase-change memory cell. For an optical programming, the write and erase pulses are sent to partially crystallize and amorphize PCM while the electrical resistance and optical transmission of the device can be monitored. In the absence of a pulse, the LR-DLSPP waveguide is in the maximum transmission regime where the mode effective index below metal stripe is close to the mode effective index above metal stripe. Thus, the PCM is in the amorphous state that results in an increased electrical resistance. However, under a write pulse the PCM is heated and crystallizes. In consequence, the equilibrium in mode effective indices above and below metal stripe is disturbed and plasmonic waveguide moves in high losses regime with a minimum transmission of light as the absorption in metal stripe arise. Simultaneously, in crystallize state of PCM the electrical resistance decreases. For an erase pulse, the PCM moves back to the amorphous state that results in a maximum transmission and increased resistance. Thus, the successful operation in electro-optic domain is possible where a change in optical transmission follows the electrical switching of the device. The PCM-based waveguides offer non-volatile weights that can be reconfigured with optical or electrical signals. This process of setting the weights is also reversible, which limits the need to read from and write to electronic memories with (digital-to-analogue conversions) DACs and (analogue-to-digital conversions) ADCs [10, 16].

The proposed arrangement can be organized into large arrays and stacked three-dimensionally what boosts a processing functionality. By increasing the depth of a neural network (number of layers) the performance can be highly enhanced. In 3D networks each individual layer could serve different functionalities such as sensors, data processing or storage. However, it requires an efficient data flow between each layer.

*b. RRAM*

Proposed plasmonic configuration can be used as well with both RRAM and ferroelectric memristors. For RRAM memristors, the switching can be realized in an amorphous silicon layer placed between doped Si and metal stripe [45]. The p+Si layer is used here as an electrode and as a medium that supports the propagating LR-DLSPP plasmonic signal. In a contrary, the amorphous silicon is placed in the electric field maximum of the propagating mode, thus any changes in it cause a huge change in a propagating plasmonic mode. The electrically triggered creation of a metallic filament determines the variation of the absorption loss of the fundamental LR-DLSPP plasmonic mode – the balance between a mode effective index below metal stripe and above metal stripe is disturbed and the absorption losses arise. This results in the optical detection of an electrical RRAM switching and enables an optical readout of the memory state.

*c. Ferroelectrics*

Ferroelectrics are very interesting materials as they feature spontaneous electric polarization which can be reversibly switched by applying an external electric field [51, 52, 53, 60] or by interaction with a light [50, 55]. Optically-controlled polarization in ferroelectrics can be used to store and retrieve light information, thus the optoelectronic ferroelectric memories can be achieved [55].

The ferroelectric semiconductor based two-terminal devices are very interesting platform for nonvolatile memory and neuromorphic computing applications. They can be implemented in planar and vertical structures thus providing an additional functionality. The capability to program a memristors at both planar and vertical orientations, i.e., multidirectional programming, enables the tunning of device conducting paths through a third terminal, which is intriguing and favorable for integrating device array with optimized performances and small device-to-device variations [53].

The crossbar ferroelectric semiconductor junction (c-FSJ) devices are very promising for high-density and low-power applications [54]. To enhance the on/off ration of the FTJ, the asymmetric electrodes arrangement is commonly applied [54].

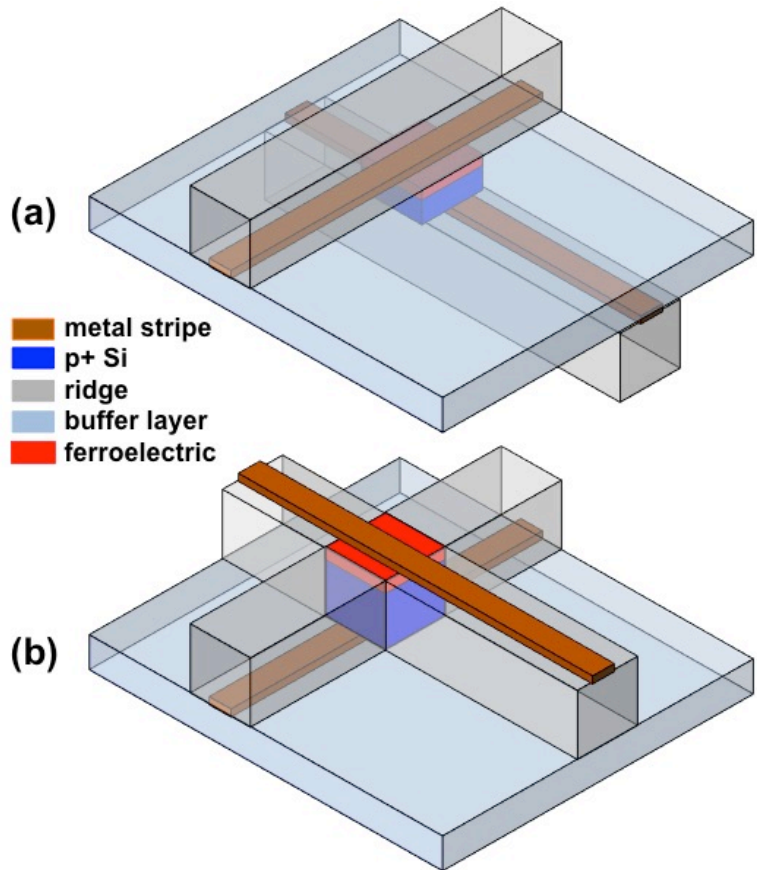

**Figure 10**. Plasmonic waveguide ferroelectric memristor arrangement in (a) "inverse" and (b) "normal" designs.

A c-FSJ can be easily implemented in proposed plasmonic crossbar arrays where, for example, a heavily boron-doped silicon (p+Si) can be used as one of the electrodes while the metal stripe supporting a propagating plasmonic mode can be used as a second electrode (Fig. 10). A depletion region in doped Si provides essential asymmetry to the band alignments in the metal/α-$In_2Se_3$/Si structure, which can enhance the modulation of the effective Schottky barrier height [54]. As it was shown in Ref. [54] in such an arrangement a high on/off ratio exceeding $10^4$ can be achieved at room temperature.

It should be emphasized here that p+Si can be replaced by any other metal electrode, any other conductive material or insulator as presented in Ref. 54. Simultaneously, α-$In_2Se_3$ can be replaced by other ferroelectric material.

*Integration with modulators, photodetectors and etc.*
In terms of plasmonic waveguide, LR-DLSPP waveguide can be made with any metals that are characterized by low absorption losses. Apart from common plasmonic metals such as gold (Au), silver (Ag), aluminum (Al) or copper (Cu), we can consider as well CMOS-compatible transition metal nitrides such as for example titanium nitride (TiN) or zirconium nitride (ZrN). It will provide an additional freedom in integration with other passive and active on-chop components.

As it has been previously shown, the same LR-DLSPP waveguide platform can be used as a building block for modulators [34], switches [33] and photodetectors [27, 28, 29, 30] that can be easily integrated with ANN.

## Conclusions

We proposed a nonvolatile electro-optic crossbar array that enables both electrical and optical programming and readout using plasmonic LR-DLSPP waveguide and memory-holding materials such as PCMs, RAM, ferroelectrics or graphene. It is based on the integrated, reversible and nonvolatile memory cell that bridges the gap between electro-optic mixed-mode operations. It was enabled by implementation of plasmonic that brings electronics and photonics in one component. In consequence, both optical and electrical read and write operations were possible in a single device.

## Author information
**Affiliations**
Independent Researcher, 90-132 Lodz, Poland
Jacek Gosciniak
**Contributions**
J.G. conceived the idea, performed all calculations and FEM and FDTD simulations and wrote the article.
**Corresponding author**
Correspondence to Jacek Gosciniak (jeckug10@yahoo.com.sg)

## REFERENCES

1. J. Leuthold, C. Hoessbacher, S. Muehlbrandt, A. Melikyan, M. Kohl, C. Koos, W. Freude, V. Dolores-Calzadilla, M. Smit, I. Suarez, J. Martinez-Pastor, E. P. Fitrakis and I. Tomkos, "Plasmonics Communications: Light on a Wire," 2013, 24(5), 28-35.
2. D. V. Christensen et al., "2021 Roadmap on Neuromorphic Computing and Engineering," arXiv **2021**, arXiv:2105.05956.
3. A. Mehonic, A. Sebastian, B. Rajendran, O. Simeone, E. Vasilaki, and A. J. Kenyon, "Memristors—From In-Memory Computing, Deep Learning Acceleration, and Spiking Neural Networks to the Future of Neuromorphic and Bio-Inspired Computing, "Adv. Intell. Syst. **2020**, 2, 2000085.
4. J. Woo, J. H. Kim, J.-P. Im, and S. E. Moon, "Recent Advancements in Emerging Neuromorphic Device Technologies," Adv. Intell. Syst. **2020**, 2000111.
5. V. K. Sangwan and M. C. Hersam, "Neuromorphic nanoelectronic materials," Nat. Nanotechnology **2020**, 15, 517-528.
6. N. Farmakidis, N. Youngblood, X. Li, J. Tan, H. L. Swett, Z. Cheng, C. D. Wright, W. H. P. Pernice and H. Bhaskaran, „Plasmonic nanogap enhanced phase-change devices with dual electrical-optical functionality," Sci. Adv. **2019**, 5:eaaw2687.
7. E. Gemo, J. Faneca, S. G.-C. Carrillo, A. Baldycheva, W. H. P. Pernice, H. Bhaskaran and C. D. Wright, „A plasmonically enhanced route to faster and more energy-efficient phase-change integrated photonic memory and computing devices," Appl. Phys. **2021**, 129, 110902.
8. S. Abdollahramezani, O. Hemmatyar, M. Taghinejad, H. Taghinejad, A. Krasnok, A. A. Eftekhar, Ch. Teichrib, S. Deshmukh, M. El-Sayed, E. Pop, M. Wuttig, A. Alu, W. Cai and A. Adibi, "Electrically driven programmable phase-change meta-switch reaching 80% efficiency," arxiv **2021**, arxiv:2104.10381.
9. J. Zheng, Z. Fang, Ch. Wu, S. Zhu, P. Xu, J. K. Doylend, S. Deshmukh, E. Pop, S. Dunham, M. Li and A. Majumdar, "Nonvolatile Electrically Reconfigurable

Integrated Photonic Switch Enabled by a Silicon PIN Diode Heater," Adv. Mater. **2020**, 2001218.
10. Q. Xia and J. J. Yang, "Memristive crossbar arrays for brain-inspired computing," Nat. Materials **2019**, 18, 309–323.
11. B. J. Shastri, A. N. Tait, T. F. de Lima, M. A. Nahmias, H. T. Peng and P. R. Prucnal, "Neuromorphic photonics, principles of," Encyclopedia of Complexity and Systems Science **2018**, 1-37.
12. P. R. Prucnal, B. J. Shastri, A. N. Tait, M. A. Nahmias, T. Ferreira de Lima, "Neuromorphic photonics," CRC Press **2017**, Boca Raton, FL, USA.
13. M.-K. Kim, Y. Park, I.-J. Kim, and J.-S. Lee, "Emerging Materials for Neuromorphic Devices and Systems," iScience **2020**, 23, 101846.
14. Y. Zhang, J. B. Chou, J. Li, H. Li, Q. Du, A. Yadav, S. Zhou, M. Y. Shalaginov, Z. Fang, H. Zhong, Ch. Roberts, P. Robinson, B. Bohlin, C. Rios, H. Lin, M. Kang, T. Gu, J. Warner, V. Liberman, K. Richardson and J. Hu, "Broadband transparent optical phase change materials for high-performance nonvolatile photonics, " Nat. Comm. **2019**, 10, 4279.
15. J. Zhang, J. Zheng, P. Xu, Y. Wang and A. Majumdar, "Ultra-low-power nonvolatile integrated photonic switches and modulators based on nanogap-enhanced phase-change waveguides," Opt. Express **2020**, 28(5) 37265-37275.
16. Ch. Wu, H. Yu, S. Lee, R. Peng, I. Takeuchi and M. Li, "Programmable phase-change metasurfaces on waveguides for multimode photonic convolutional neural network," Nat. Comm. **2021**, 12, 96.
17. M. Wuttig, H. Bhaskaran and T.Taubner, "Phase-change materials for non-volatile photonic applications," Nat. Photonics **2017**, 11, 465-476.
18. K. Kato, M. Kuwahara, H. Kawashima, T. Tsuruoka and H. Tsuda, "Current-driven phase-change optical gate switch using indium–tin-oxide heater," Appl. Phys. Express **2017**, 10, 072201.
19. T. Ferreira de Lima, B. J. Shastri, A. N. Tait, M. A. Nahmias and P. R. Prucnal, "Progress in neuromorphic photonics," Nanophotonics **2017**, 6(3), 577-599.
20. Z. Cheng, C. Ríos, W. H. P. Pernice, C. D. Wright, H. Bhaskaran, "On-chip photonic synapse," Sci. Adv. **2017**;3: e1700160.
21. E. Goi, Q. Zhang, X. Chen, H. Luan and M. Gu, "Perspective on photonic memristive neuromorphic computing," PhotoniX **2020**, 1(1), 1-26.
22. A. Lugnan, A. Katumba, F. Laporte, M. Freiberger, S. Sackesyn, C. Ma, E. Gooskens, J. Dambre and P. Bienstman, "Photonic neuromorphic information processing and reservoir computing," APL Photon. **2020**, 5, 020901.
23. B. J. Shastri, A. N. Tait, T. Ferreira de Lima, W. H. P. Pernice, H. Bhaskaran, C. D. Wright and P. R. Prucnal, "Photonics for artificial intelligence and neuromorphic computing," Nat. Photonics **2021**, 15, 102-114.
24. P. Stark, F. Horst, R. Dangel, J. Weiss and B. Jan Offrein, "Opportunities for integrated photonic neural networks," Nanophotonics **2020**, 9(13), 4221-4232.
25. J. Gosciniak, T. Holmgaard and S. I. Bozhevolnyi, "Theoretical analysic of long-range dielectric-loaded surface plasmon polariton waveguides," J. of Lightw. Technol. **2011**, 29(10), 1473-1481.
26. V. S. Volkov, Z. Han, M. G. Nielsen, K. Leosson, H. Keshmiri, J. Gosciniak, O. Albrektsen and S. I. Bozhevolnyi, "Long-range dielectric-loaded surface plasmon polariton waveguides operating at telecommunication

wavelengths," Opt. Lett. **2011**, 36 (21), 4278-4280.
27. J. Gosciniak, and J. B. Khurgin, "On-Chip Ultrafast Plasmonic Graphene Hot Electron Bolometric Photodetector," ACS Omega **2020**, 5 (24), 14711-14719.
28. J. Gosciniak, F. B. Atar, B. Corbett, and M. Rasras, "Plasmonic Schottky photodetector with metal stripe embedded into semiconductor and with a CMOS-compatible titanium nitride," Sci. Rep. **2019**, 9 (1), 6048.
29. J. Gosciniak and M. Rasras, "High-bandwidth and high-responsivity waveguide-integrated plasmonic germanium photodetector," JOSA B **2019**, 36 (9), 2481-2491.
30. J. Gosciniak, M. Rasras and J. Khurgin, "Ultrafast Plasmonic Graphene Photodetector Based on the Channel Photothermoelectric Effect," ACS Photonics **2020**, 7 (2), 488-498.
31. J. Gosciniak, M. G. Nielsen, L. Markey, A. Dereux and S. I. Bozhevolnyi, "Power monitoring in dielectric-loaded plasmonic waveguides with internal Wheatstone bridges," Opt. Exp. **2013**, 21 (5), 5300-5308.
32. J. Gosciniak and S. I. Bozhevolnyi, "Performance of thermo-optic components based on dielectric-loaded surface plasmon polariton waveguides," Sci. Rep. **2013**, 3 (1), 1-8.
33. J. Gosciniak, L. Markey, A. Dereux and S. I. Bozhevolnyi, "Themo-optic control of dielectric-loaded plasmonic Mach-Zehnder interferometers and directional coupler switches," Nanotechnology **2012**, 23 (44), 444008.
34. J. Gosciniak, D. T. H. Tan, "Theoretical investigation of graphene-based photonic modulators," Sci. Rep. **2013**, 3, 1897.
35. S. G. Kim, J. S. Han, H. Kim, S. Y. Kim and H. Wo. Jang, "Recent Advances in Memristive Materials for Artificial Synapses," Adv. Mat. Technologies **2018**, 3 (12), 1800457.
36. G. Di Martino and S. Tappertzhofena, "Optically accessible memristive devices," Nanophotonics **2019**, 8(10), 1579–1589.
37. Z. Liu, J. Tang, B. Gao, X. Li, P. Yao, Y. Lin, D. Liu, B. Hong, H. Qian and H. Wu, "Multichannel parallel processing of neural signals in memristor arrays," Sci. Adv. **2020**, 6, eabc4797.
38. Z. Liu, J. Tang, B. Gao, P. Yao, X. Li, D. Liu, Y. Zhou, H. Qian, B. Hong and H. Wu, "Neural signal analysis with memristor arrays towards high-efficiency brain–machine interfaces," Nat. Comm. **2020**, 11, 4234.
39. L. Chua, "Memristor – The missing circuit element," IEEE Trans. Circuit Theory **1971**, 18, 507-519.
40. D. B. Strukov, G. S. Snider, D. R. Stewart and R. S. Williams, "The missing memristor found," Nature **2008**, 453, 80-83.
41. WO2012/177265 12/27/2012 Miao et al. "High-Reliability High-Speed Memristor"
42. 2012/0014170 A1 1/19/2012 Strukov et al. "Capacitive Crossbar Arrays"
43. 2017/0358352 A1 12/14/2017 Ge et al. "Nonvolatile Memory Cross-bar Array"
44. B. Cheng, A. Emboras, Y. Salamin, F. Ducry, P. Ma, Y. Fedoryshyn, S. Andermatt, M. Luisier and J. Leuthold, "Ultra compact electrochemical metallization cells offering reproducible atomic scale memristive switching," Communications Physics **2019**, 2, 28.
45. A. Emboras, I. Goykhman, B. Desiatov, N. Mazurski, L. Stern, J. Shappir and U. Levy, "Nanoscale Plasmonic Memristor with Optical Readout Functionality,"

Nano Lett. **2013**, 13, 6151–6155.
46. A. Emboras, A. Alabastri, F. Ducry, B. Cheng, Y. Salamin, P. Ma, S. Andermatt, B. Baeuerle, A. Josten, Ch. Hafner, M. Luisier, P. Nordlander, and J. Leuthold, "Atomic Scale Photodetection Enabled by a Memristive Junction," ACS Nano **2018**, 12, 6706–6713.
47. S. Raoux, F. Xiong, M. Wuttig, and E. Pop, "Phase change materials and phase change memory," MRS Bull. **2014**, 39, 703.
48. J. Feldmann, M. Stegmaier, N. Gruhler, C. Ríos, H. Bhaskaran, C.D. Wright and W.H.P. Pernice, "Calculating with light using a chip-scale all-optical abacus," Nat. Comm. **2017**, 8, 1256.
49. C. Rios, Y. Zhang, M. Y. Shalaginov, S. Deckoff-Jones, H. Wang, S. An, H. Zhang, M. Kang, K. A. Richardson, Ch. Roberts, J. B. Chou, V. Liberman, S. A. Vitale, J. Kong, T. Gu and J. Hu, "Multi-Level Electro-Thermal Switching of Optical Phase-Change Materials Using Graphene," Adv. Photonics Res. **2020**, 2000034.
50. K. Xu, W. Jiang, X. Gao, Z. Zhao, T. Lowb and W. Zhu, "Optical control of ferroelectric switching and multifunctional devices based on van der Waals ferroelectric semiconductors," Nanoscale **2020**, 12, 23488-23496.
51. N. Higashitarumizu, H. Kawamoto, Ch.-J. Lee, B.-H. Lin, F.-H. Chu, I. Yonemori, T. Nishimura, K. Wakabayashi, W.-H. Chang and K. Nagashio, "Purely in-plane ferroelectricity in monolayer SnS at room temperature," Nat. Comm. **2020**, 11, 2428.
52. L. Wang, X. Wang, Y. Zhang, R. Li, T. Ma, K. Leng, Z. Chen, I. Abdelwahab and K. P. Loh, "Exploring Ferroelectric Switching in α-$In_2Se_3$ for Neuromorphic Computing," Adv. Funct. Mater. **2020**, 2004609.
53. F. Xue, X. He, J. R. D. Retamal, A. Han, J. Zhang, Z. Liu, J.-K. Huang, W. Hu, V. Tung, J.-H. He, L.-J. Li and X. Zhang, "Gate-Tunable and Multidirection-Switchabe Memristive Phenomena in a Van Der Waals Ferroelectric," Adv. Mater. **2019**, 1901300.
54. M. Si, Z. Zhang, S.-Ch. Chang, N. Haratipour, D. Zheng, J. Li, U. E. Avci and P. D. Ye, "Asymmetric Metal/α-In2Se3/Si Crossbar Ferroelectric Semiconductor Junction," ACS Nano **2021**, 15, 3, 5689-5695.
55. F. Xue, X. He, W. Liu, D. Periyanagounder, Ch. Zhang, M. Chen, Ch.-H. Lin, L. Luo, E. Yengel, V. Tung, T. D. Anthopoulos, L.-J. Li, J,-H. He, and X. Zhang, "Optoelectronic Ferroelectric Domain-Wall Memories Made from a Single Van Der Waals Ferroelectric," Adv. Funct. Mater. **2020**, 2004206.
56. T.-J. Ko, H. Li, S. A. Mofid, Ch. Yoo, E. Okogbue, S. S. Han, M. S. Shawkat, A. Krishnaprasad, M. M. Islam, D. Dev, Y. Shin, K. H. Oh, G.-H. Lee, T. Roy and Y. Jung "Two-Dimensional Near-Atom-Thickness Materials for Emerging Neuromorphic Devices and Applications," iScience **2020**, 23, 101676.
57. T. F. Schranghamer, A. Oberoi and S. Das, "Graphene memristive synapses for high precision neuromorphic computing," Nat. Comm. **2020**, 11, 5474.
58. A. J. Arnold, A. Razavieh, J. R. Nasr, D. S. Schulman, Ch. M. Eichfeld and S. Das, "Mimicking Neurotransmitter Release in Chemical Synapses via Hysteresis Engineering in $MoS_2$ Transistors," ACS Nano **2017**, 11, 3110–3118.
59. M. Si, A. K. Saha, S. Gao, G. Qui, J. Qin, Y. Duan, J. Jian, Ch. Niu, H. Wang, W. Wu, S. K. Gupta and P. D. Ye, "A ferroelectric semiconductor field-effect transistor," Nat. Electronics **2019**, 2, 580-586.
60. X. Chai, J. Jiang, Q. Zhang, X. Hou, F. Meng, J. Wang. L. Gu, D. W. Zhang and A. Q.

Jiang, "Nonvolatile ferroelectric field-effect transistors," Nat. Comm. **2020**, 11, 2811.
61. Ch.-Y. Wang, C. Wang, F. Meng, P. Wang, S. Wang, S.-J. Liang and F. Miao, "2D Layered Materials for Memristive and Neuromorphic Applications," Adv. Electronic Materials **2020**, 6 (2), 1901107.
62. Q. Zhang, Y. Zhang, J. Li, R. Soref, T. Gu and J. Hu, "Broadband nonvolatile photonic switching based on optical phase change materials: beyond the classical figure-of-merit," Opt. Lett. **2018**, 43 (1), 94-97.
63. M. Delaney, I. Zeimpekis, D. Lawson, D. W. Hewak and O. L. Muskens, "A New Family of Ultralow Loss Reversible Phase-Change Materials for Photonic Integrated Circuits: Sb2S3 and Sb2Se3," Adv. Funct. Mater. **2020**, 2002447.